\documentclass[12pt]{article}

\begin{document}

\vspace*{2.0cm}

\begin{center}
{\large {\bf Systematic Study of Triaxial Deformation \\
             in the Relativistic Mean Field Theory\\ }  }
\vspace*{1.0cm}
S. Sugimoto$^{a,b,}$\footnote{e-mail: satoru@postman.riken.go.jp},
K. Sumiyoshi$^{a,}$\footnote{e-mail: sumi@postman.riken.go.jp},
D. Hirata$^{a,c,}$\footnote{e-mail: daisy@sp8sun.spring8.or.jp},
B.V. Carlson$^{d,}$\footnote{e-mail: brett@fis.ita.cta.br }, \\
I. Tanihata$^{a,}$\footnote{e-mail: tanihata@rikaxp.riken.go.jp},
and H. Toki$^{a,b,}$\footnote{e-mail: toki@rcnp.osaka-u.ac.jp} \\
\vspace*{0.5cm}
      $^{a}$The Institute of Physical and Chemical Research (RIKEN), \\
            Hirosawa, Wako, Saitama 351-0198, Japan \\
      $^{b}$Research Center for Nuclear Physics(RCNP), Osaka University, \\
            Mihogaoka, Ibaraki, Osaka 567-0047, Japan \\
      $^{c}$Japan Synchrotron Radiation Research Institute (JASRI)
            - SPring-8, \\
            Sayo, Hyogo, 679-5198, Japan \\
      $^{d}$Departamento de F\'{\i}sica,
            Instituto Tecnol\'ogico da Aeron\'autica - CTA, \\
            S\~ao Jos\'e dos Campos, S\~ao Paulo, Brazil
\end{center}

\vspace*{0.5cm}
\newpage
\begin{abstract}
We use the relativistic mean field (RMF) theory to systematically study
the change of deformation of
even-even nuclei in the proton-rich Xe region.
We investigate the appearance of triaxial deformation
in 25 nuclei
in the region covering $Z=50-58$ and $N=64-72$
by performing constrained, triaxially symmetric RMF calculations of their
energy surfaces.
We include pairing correlations using the BCS formalism.
We find that the Sn isotopes are spherical and the Te isotopes are very gamma unstable
with shallow minima around $\gamma = 60^\circ$.
Adding more protons, the Xe , Ba and Ce isotopes have prolate deformations with their sizes
increasing with proton number.
The neutron number dependence is found to be small.
We compare the calculated results with the available experimented data on the binding energy and the radii.
\end{abstract}


\newpage

\section{Introduction}

The recent progress of radioactive nuclear beam facilities
has provided us with marvelous findings in nuclear
physics.
Exotic structures such as neutron halos \cite{Tan85}
and neutron skins \cite{Tan92,Suz95}
have been found in experimental studies of light unstable
nuclei in the neutron-rich region.
Much new information on the shapes and structures of nuclei far from
stability is being revealed by the systematic measurement
of radii and moments of unstable nuclei \cite{Ham89,Ott89}.
Planned facilities in the world will access
a large number of unstable nuclei in the whole region
of the nuclear chart and enable us to explore
where and how exotic phenomena of nuclear structure
appear in the region far from the stability line
\cite{Tan95,RNB4}.
One of the great interests is to know where the deformation
of unstable nuclei appears and how the shape of
these nuclei changes along the isotopic and isotonic chains.

At the same time, the relativistic many-body framework has
been extensively applied to study nuclei and
nuclear matter \cite{Ser86,Ser92}.
This has been motivated by the recent success of the relativistic
Brueckner-Hartree-Fock (RBHF) theory, in which the
strong density-dependent repulsion arises automatically
from the relativistic many-body treatment, in reproducing
the saturation property of nuclear matter
\cite{Bro90,Eng94,Bro92}.
Among other properties, the relativistic mean field (RMF) theory,
which is the phenomenological framework of the RBHF theory,
has been shown to be excellent at describing the properties
of unstable nuclei as well as stable ones
\cite{Rei86,Gam90,Hir91,Sum93,Pom97}.
The RMF theory has also been successfully applied to the study
of the deformation of nuclei, as well as other
properties of the stable and unstable nuclei
\cite{Sha92,Hir93,Lal95}.
Furthermore, the RMF theory has been used to calculate the equation of state(EOS) of nuclear matter in the wide
density and temperature regions tabulated for the application to supernova simulations \cite{Shen98}.
Recently, a systematic study of all even-even
nuclei up to the drip lines in the nuclear chart
has been performed in the RMF theory with axial
deformation \cite{Hir97}.
The ground state properties of about 2000
even-even nuclei from $Z=8$ to $Z=120$ have
been studied and all possible deformations
of each nuclide have been surveyed using a
constrained, axially deformed RMF model.
Through the systematic analysis of the ground state
deformations thus found, the pattern of
the appearance of prolate and oblate deformations
has been obtained. In the same study,
it was also found that
the coexistence of prolate and oblate shapes
with similar binding energies occurs
in many nuclei in the nuclear chart.
This coexistence suggests the possible appearance of
deformation beyond the axial kind, such as
triaxial or even higher order multipole deformations.

The appearance of triaxial deformation
in this context has been studied in the case
of the neutron-rich Sulfur isotopes \cite{Hir96}.
In the axial RMF calculation for neutron-rich
Sulfur isotopes, the energy curves as a function
of the $\beta$ deformation have two minima,
at both prolate and oblate deformations, with energies
very close to each other.
Judging solely from the energy curve, one cannot
conclude which ground state deformation is
realized or whether yet another type of
deformation appears.
The RMF calculations with triaxial deformation
of the same isotopes have been performed to clarify
this point and a smooth shape transition from
prolate to oblate shapes through triaxial shape
has been found along the Sulfur isotopic chain \cite{Hir96}.
This example motivates us to study further
the appearance of triaxial deformation
in other regions of the nuclear chart.
It is interesting to explore where
triaxial deformation appears in the nuclear
chart, especially in relation with the behavior
of the appearance of axial deformation.

In the present study, we have chosen to explore the proton-rich
Xe region for the appearance of triaxial deformation.
We have made a systematic study of 25 even-even
nuclei covering $Z=50-58$ and $N=64-72$, using the
RMF theory with triaxial deformation,
in order to clarify how their shapes change as a
function of $N$ and $Z$ in this region.
We have calculated the energy surface of those nuclei
as a function of the deformation parameters,
$\beta$ and $\gamma$, to explore the ground state
deformation.
The previous study of $_{54}$Xe, $_{55}$Cs and $_{56}$Ba
isotopes using the RMF theory with axial deformation \cite{Hir95}
was successful in reproducing the general features of
the ground state properties.
However, disagreement
with the measured isotope shift for the proton-rich region, which
might be due to triaxial deformation, was observed.
In the systematic RMF calculation with axial
deformation \cite{Hir97}, which we will discuss in Sect. 3,
the shape change from oblate to prolate shape occurs as
$Z$ increases, in a region in which the two shapes coexist.
Thus, the axially symmetric RMF calculations strongly suggest
that this region could contain
triaxial deformed nuclei.

This region has been discussed as a possible
region for triaxial deformation in studies using conventional
frameworks \cite{Cas85,Wys89,Gel93,Mey97}.
There have also been experimental efforts to measure
excitation energies in order to study the collective nature
of the nuclei in this region \cite{Dew92,Pet94}.
Studies of triaxial deformation within the
mean field approach have also been performed in other
regions of the nuclear chart
\cite{Bon85,Koe88,Abe90,Wer94}.

In Sect. 2, we describe the framework of the
RMF theory with deformation.
We discuss the behavior of the shape within the RMF theory
under the assumption of axial symmetry in Sect. 3.
We present the results of the calculations
in the RMF theory with triaxial deformation in
Sect. 4.  Results of the calculations are discussed
in Sect. 5.  We summarize the paper in Sect. 6.

\section{Relativistic mean field theory}

We briefly describe the framework of the RMF theory
and the procedure of the calculation.
All details can be found in \cite{Ser86,Gam90,Hir96}.
In the RMF theory, the system of nucleons is described by fields
of mesons and nucleons under the mean field approximation.
We start with the effective lagrangian, which is relativistically covariant,
composed of meson and nucleon fields.
We adopt a lagrangian with
non-linear  $\sigma$ and $\omega$ terms \cite{Sug94},
\begin{eqnarray}
{\cal L}_{RMF} & = & \bar{\psi}\left[i\gamma_{\mu}\partial^{\mu} -M
-g_{\sigma}\sigma-g_{\omega}\gamma_{\mu}\omega^{\mu}
-g_{\rho}\gamma_{\mu}\tau_a\rho^{a\mu}
-e\gamma_{\mu}\frac{1-\tau_3}{2}A^{\mu}\right]\psi  \\ \nonumber
 && +\frac{1}{2}\partial_{\mu}\sigma\partial^{\mu}\sigma
-\frac{1}{2}m^2_{\sigma}\sigma^2-\frac{1}{3}g_{2}\sigma^{3}
-\frac{1}{4}g_{3}\sigma^{4} \\ \nonumber
 && -\frac{1}{4}W_{\mu\nu}W^{\mu\nu}
+\frac{1}{2}m^2_{\omega}\omega_{\mu}\omega^{\mu}
+\frac{1}{4}c_{3}\left(\omega_{\mu}\omega^{\mu}\right)^2   \\ \nonumber
 && -\frac{1}{4}R^a_{\mu\nu}R^{a\mu\nu}
+\frac{1}{2}m^2_{\rho}\rho^a_{\mu}\rho^{a\mu}
-\frac{1}{4}F_{\mu\nu}F^{\mu\nu} ,
\end{eqnarray}
where the notation follows the standard one.  On top of the Walecka
$\sigma$ - $\omega$ model with photons and isovector-vector $\rho$ mesons,
non-linear $\sigma$ meson terms are introduced to
reproduce the properties of nuclei quantitatively and give a
reasonable value for the incompressibility \cite{Bog77}.  The inclusion of the
non-linear term of $\omega$ meson \cite{Bod91}
is motivated by the recent success of the
relativistic Brueckner-Hartree-Fock theory \cite{Bro90,Sug94}.  Deriving the
Euler-Lagrange equations from the lagrangian under the mean field
approximation, we obtain the Dirac equation for the nucleons and
Klein-Gordon equations for the mesons.  The self-consistent Dirac
equation and Klein-Gordon equations are solved by expanding the fields in
terms
of harmonic-oscillator wave functions \cite{Gam90,Hir96}.

The RMF model contains the meson masses, the meson-nucleon coupling
constants and the meson self-coupling constants as free parameters.  We adopt
the parameter set TMA, which was determined by fitting the experimental
data of masses and charge radii of nuclei in a wide mass range
\cite{Hir97,Sug95,Tok95}.
The parameters are listed in Table 1.
We remark that this parameter set has a mass dependence
so as to reproduce nuclear properties quantitatively
from the light mass region to the superheavy region.
With the TMA parameter set, the symmetry energy is 30.68 MeV and
the incompressibility is 318 MeV.
Note that the bulk properties of nuclear matter at saturation
with the parameter set TMA is calculated for uniform matter
in the limit of infinite mass number.

The TMA parameter set has been used for the systematic study
of all even-even nuclei up to the drip lines in the nuclear
chart within the RMF framework under the assumption of
axial symmetry \cite{Hir97}.
It has been shown that the overall agreement of the calculated
results using TMA with the experimental data of masses and charge
radii is excellent and is found to be much better than the results
of spherical RMF calculations with TMA.
In the present study, we extend the RMF calculation with the TMA
parameter set from the axially symmetric to the triaxially symmetric one,
in order to explore the appearance of triaxial deformation in the region
in which axial deformations with similar binding energies coexist.
We discuss the correspondence between the axial and triaxial RMF
calculations in the subsequent section.

In order to take into account triaxial deformation,
the fields are expanded
in terms of the eigenfunctions of a triaxially deformed harmonic oscillator
potential \cite{Hir96}.
We perform calculations that constrain the quadrupole moments
of the nucleon distribution,
 in order to survey the coexistence of multiple shapes
and to identify the ground state deformation.
We use a quadratic constraint to calculate a complete map of the energy
surface as a function of the deformation \cite{Hir96,Flo73}.
We take a basis of up to $N=12$ major shells of the harmonic oscillator
wave functions.
This is normally enough for the constrained calculations
in this mass range \cite{Hir96}.
We have performed calculations with $N=14$ major shells in the case of no-pairing
and found that the energy surface does not change significantly for moderate deformation.
We have also performed non-constrained calculations with pairing up to $N=16$
and the convergence was generally good with $N=12$.

We take the pairing window as given by Gambhir et al. \cite{Gam90} as 
$\varepsilon-\lambda \le 2(41A^{-1/3})$MeV.
As for the pairing correlations, we perform the RMF calculations with triaxial symmetry
using a BCS formalism \cite{Gam90}.
Since we calculate the energy surface in the full $\beta-\gamma$ range, 
we take the  pairing interaction strength for a given nucleus as $G=23/A$[MeV]
for both protons and neutrons \cite{Pair}.

Although the BCS-type treatment has often been used in the
RMF calculations, it would be preferable to incorporate
the pairing correlations in the relativistic many body framework
in a consistent manner.
A study with the relativistic Hartree-Bogoliubov theory
has been performed under the assumption of spherical symmetry \cite{Men96}.
A systematic study of nuclear deformation within such
a relativistic many-body framework is
currently being made \cite{Bre98}.

\section{Numerical results}

Before we present the results of the RMF calculation
with triaxial symmetry, we discuss the corresponding
results of the RMF calculation with axial symmetry
\cite{Hir97}.
We examine the behavior of the calculated ground state
properties of even-even nuclei in the proton-rich Xe region.
We note here that the calculated properties of deformations
as well as masses and radii in this region agree very well
with the experimental data.

In the region of $Z=50-58$ and $N=64-72$,
roughly speaking, the deformation changes according
to the proton number, with a few exceptions. The
Sn isotopes ($Z=50$) are dominated by a spherical
shape and the
Te isotopes ($Z=52$) have, for the most part, an oblate shape.
The Xe, Ba and Ce isotopes ($Z=54-58$) all have a prolate
shape.
We show the energy surfaces for Te and Xe isotopes as functions of $\beta$ deformation (Fig. 1).
The general trend shows a transition in
shape from spherical, through oblate, to prolate
as the proton number increases.

Furthermore, there is always more than one energy minimum
and shape coexistence is generally observed \cite{Hir97}
in this region of nuclides, as can be seen in Fig. 1.
For each nucleus there is a corresponding second minimum
which has the deformation parameter, $\beta$,
of opposite sign to that of the absolute minimum.
The energy difference between
the absolute minimum and the second minimum
is generally small in this region.
%
%

We next present the  RMF calculations with triaxial
deformation for 25 nuclei covering the range $Z=50-58$
and $N=64-72$.  We present the energy surfaces of these nuclei
in the  $\beta$ and $\gamma$ deformation parameter space,
calculated using constraints on the two deformation parameters.
Figure 2 displays the
energy surfaces of the Sn, Te, Xe, Ba, Ce ($Z=50-58$)
isotopes with $N=64-72$, arranged in the form of the
nuclear chart.
The spacing of contours
is 1 MeV in total energy in all figures.
The energy minimum is marked by the black region,
in which the energy difference is less than 1 MeV from its absolute minimum energy.

As for the Sn isotopes, the minima appear consistently at the spherical shape ($\beta=0$). 
The Te isotopes show a $\gamma$-soft character,
having similar energies along the $\gamma$ direction.
Most of the Xe, Ba and Ce isotopes have
minima at prolate deformation.
In some cases ($^{124}$Ba, $^{128}$Ba, $^{126}$Ce,  $^{128}$Ce and $^{130}$Ce),
the region of the shallow minimum
extends to quite large $\gamma$ deformation.
These results indicate that the triaxial shape is not stable in those nuclei,
but the energy surfaces are very $\gamma$ soft.
\section{Discussion}

We discuss here the influence of the pairing
correlations and the effective interaction
on the triaxial RMF calculations.
Since the consistent calculation of pairing
correlations with deformation within the relativistic
many-body framework is still
under development, we have performed the
RMF calculations with pairing correlations in the BCS formalism,
as a first study of the appearance
of axial and triaxial deformation.
We see here the effect of pairing correlations on
the magnitude of the binding energy and the deformation
by comparing the results of calculations with and without pairing.
We show in Fig. 3 the energy surfaces in the $\beta-\gamma$ plane of $^{120}$Te, $^{122}$Xe
and $^{124}$Ba without and with pairing correlations.
The qualitative behavior of the two cases is very similar.
Generally speaking, the magnitude of the $\beta$ deformation is reduced by the pairing correlations.
The energy minimum seen at finite $\gamma$ deformation in $^{124}$Ba is washed out by the
inclusion of the pairing correlations.

In Fig. 4 we show the proton deformation parameters $\beta$ extracted from
the RMF calculation with axial deformation.
In the same figure we also show the experimental data obtained from the B(E2) values\cite{Ram87}.
In this figure we see that our results compare  well with experimental data.

We also calculate the energy surface of $^{124}$Ba with alternative NL1 
parameter set \cite{Rei86}
of the RMF theory in order to test the dependence of triaxial deformation on the choice of the RMF parameter.
In Fig. 5, we compare the energy surfaces obtained by using the TMA and NL1 parameter sets.
A well-distinguished minimum is seen at prolate deformation for the case of the NL1 parameter set.
This feature is slightly stronger than in the TMA case.
We mention that triaxial deformation is not found in calculations within the Skyrme-Hartree-Fock(SHF) theory\cite{Taj96}.
It would be interesting to compare the energy surfaces of RMF and SHF theories in a wide mass range.

In Fig. 6 we show the total binding energy of the nuclei studied.
The theoretical values are taken from the axially symmetric calculations,
since no distinguished triaxial shapes were found in the triaxial calculations. 
We see slightly over binding for the Sn isotopes, which may be due to the use
of the constant pairing correlations and will be studied further.
We show in Fig. 7 the charge radius as a function of the neutron number.
The general tendency is found to be quite satisfactory.
%
%
%
\section{Summary}

We have studied systematically the triaxial deformation
of 25 even-even nuclei in the proton-rich Xe region.
We have calculated their ground state structures
in the RMF theory with triaxial deformation and with pairing correlations
and obtained their energy surfaces in the plane
of the deformation parameters, $\beta$ and $\gamma$,
by constraining the quadrupole moments.
We have explored the appearance of triaxial
deformation in the region covering $Z=50-58$ and $N=64-72$,
by looking for the minima of the derived
energy surfaces in the triaxial deformation parameter space.
Through comparisons with the results obtained
in the RMF calculations with axial symmetry,
we have discussed the correspondence between the coexistence
of axial shapes and the appearance of triaxial
shapes.

We have found no distinguished energy minima at triaxial deformations. However,
the energy surfaces are often very $\gamma$ soft. This feature is caused by the
pairing correlations since, when we remove the pairing correlations in the RMF
calculations,  we find well distinguished triaxial deformation in this mass region.
We have compared the energy surfaces of two parameter sets, the TMA and NL1 ones.
The TMA parameter set provides more softness in the $\gamma$ direction than the NL1 one provides.
Comparisons of the binding energies and deformations with experimental data are, in general,
quite satisfactory.

We note here that we have not worked out the angular momentum and particle number projection,
which have already been developed in the non-relativistic approach.
The restorations of these symmetries may change somewhat the results on the deformation,
as has been discussed in the non-relativistic description of deformed nuclei.
The relativistic approach to triaxial deformation is not yet at such a level
of systematic study and these refinements have not been
considered here. This is certainly a direction for future work.
\section*{Acknowledgment}

We would like to thank J. Meng for fruitful discussions.
The entire calculation was performed on
the Fujitsu VPP500/30 and VPP700E/128 supercomputer at RIKEN, Japan.
K. S. would like to express special thanks to the Computing
Facility of RIKEN for a special allocation of VPP500/30 computing time
for the first stage of this study.


\newpage

\begin{center} {\bf Table 1}
\end{center}
\[\begin{tabular}{|c|c|} \hline
$m_{N}$ [MeV]	   &  938.900  \\ \hline
$m_{\sigma}$ [MeV] &  519.151  \\ \hline
$m_{\omega}$ [MeV] &  781.950  \\ \hline
$m_{\rho}$ [MeV]   &  768.100  \\ \hline
$g_{\sigma}$	   &   10.055	$+$    3.050/$A^{0.4}$  \\ \hline
$g_{\omega}$	   &   12.842	$+$    3.191/$A^{0.4}$  \\ \hline
$g_{\rho}$	       &    3.800	$+$    4.644/$A^{0.4}$  \\ \hline
$g_{2}$	           & $-$0.328	$-$   27.879/$A^{0.4}$  \\ \hline
$g_{3}$     	   &   38.862	$-$  184.191/$A^{0.4}$  \\ \hline
$c_{3}$	           &  151.590	$-$  378.004/$A^{0.4}$  \\ \hline
\end{tabular} \]
\newpage

\section*{Figure captions}

\begin{description}

\item[Fig.1]
The energy curve obtained in axial RMF calculations
as a function of the deformation parameter, $\beta$,
for the $_{52}$Te and $_{54}$Xe isotopes.  Calculated points
are connected by dashed curves to guide the eye.
Note that the curves for $^{122}$Te and $^{124}$Te are shifted downward by 0.02 MeV to
distinguish them from the other curves.


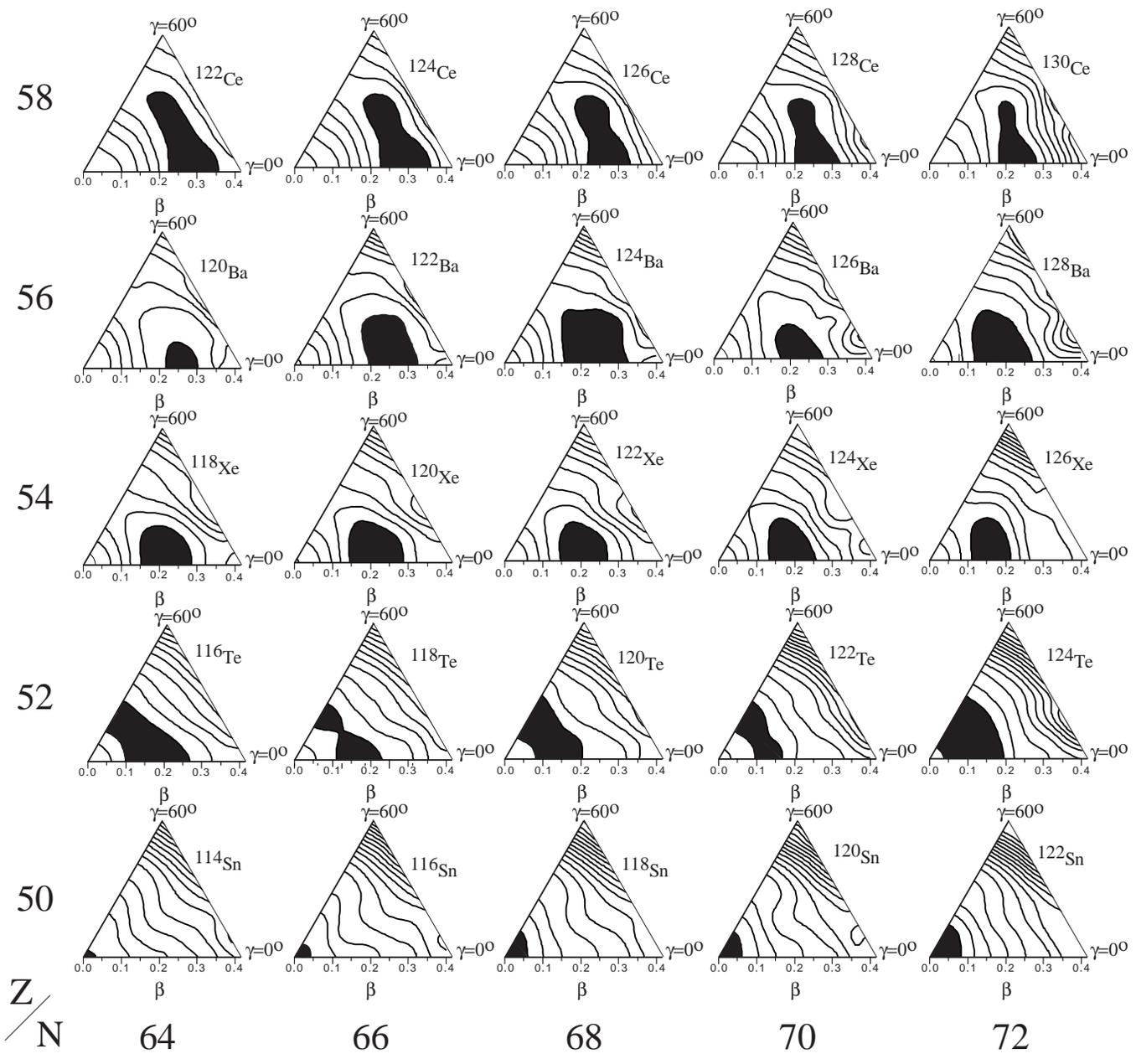
\item[Fig.2]
The energy surface in the plane of deformation parameters,
$\beta$ and $\gamma$, calculated in the RMF theory
with triaxial deformation for nuclei in the range of
$Z=50-58$ and $N=64-72$, arranged in the form of the nuclear chart.
The energy difference between the contours is 1 MeV in total binding energy.
The energy minimum is marked by the black region,
in which the energy difference is less than 1 MeV.

\item[Fig.3]
The energy surfaces obtained in the triaxial RMF calculation
without and with pairing are shown for $^{120}$Te, $^{122}$Xe and $^{124}$Ba.
The energy difference between the contours is 1 MeV in total binding energy.
The energy minimum is marked by the black region.

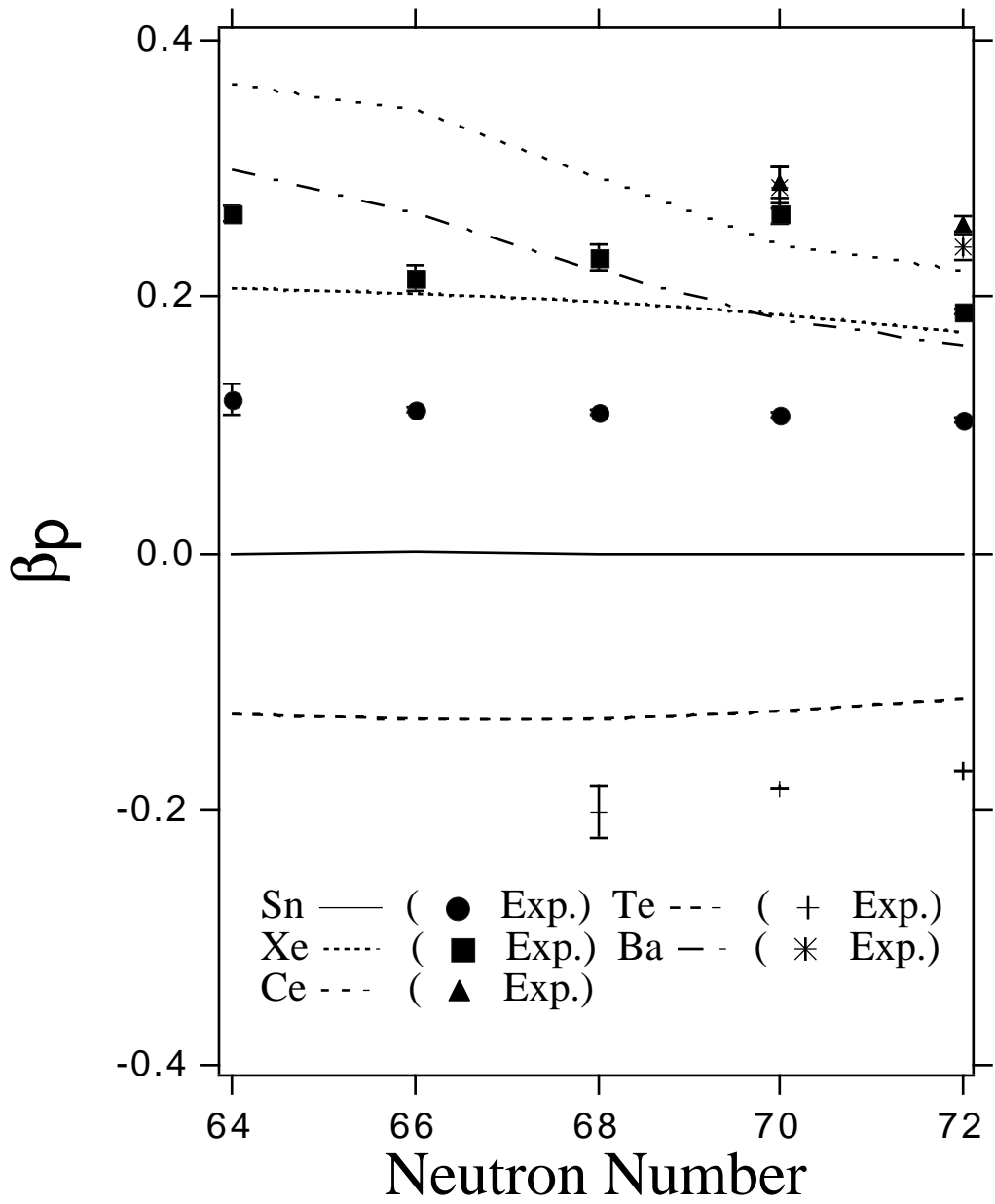
\item[Fig.4]
The proton deformation parameter $\beta$, obtained from the RMF calculation
with axial deformation, is shown as a function of the neutron number.
The experimental data are extracted from Ref.~\cite{Ram87}.

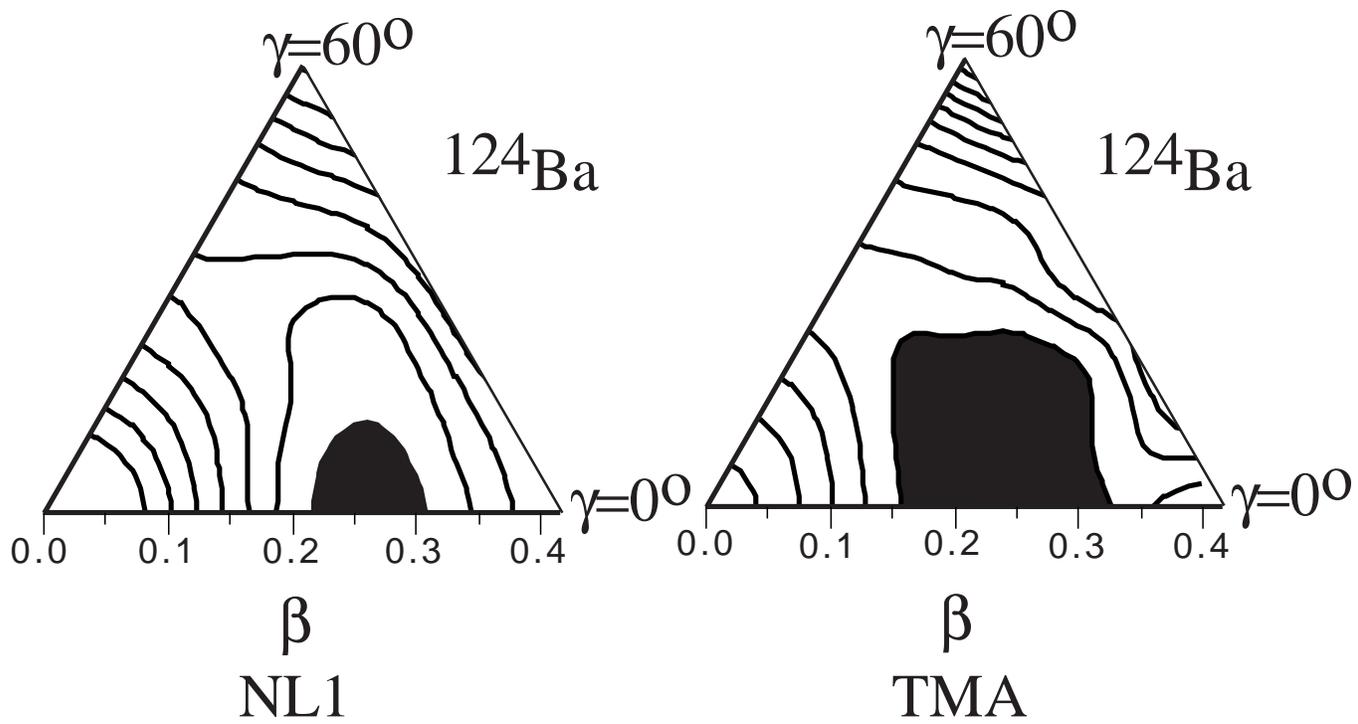
\item[Fig.5]
The energy surfaces obtained in the triaxial RMF calculation
with the NL1 and TMA parameter sets are shown for $^{124}$Ba.
The energy difference between the contours is 1 MeV in total binding energy.
The energy minimum is marked by the black region.

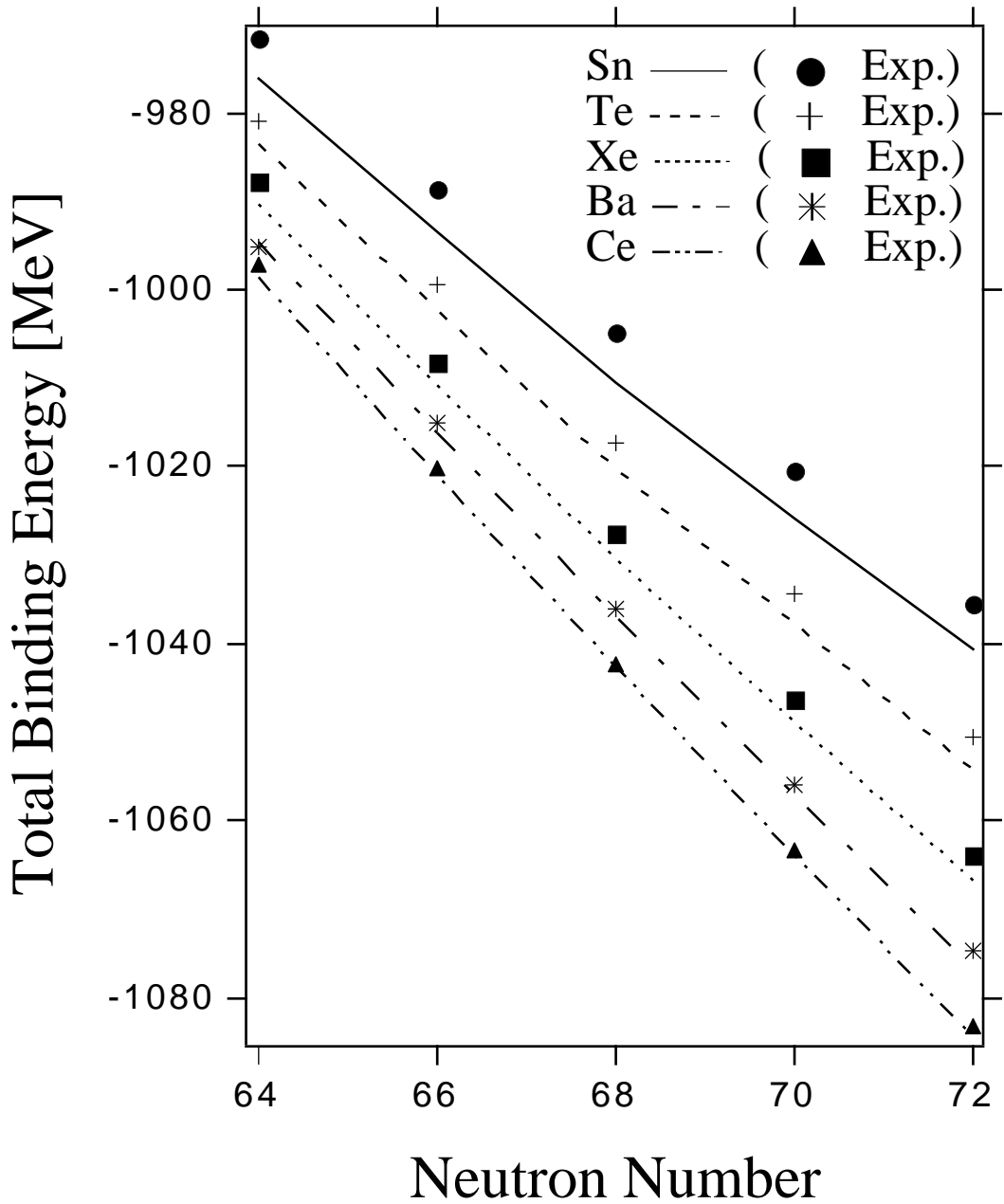
\item[Fig.6]
The total binding energy calculated using the RMF theory with axial deformation is
shown as a function of the neutron number.
The experimental data are extracted from Ref.~\cite{Aud95}.

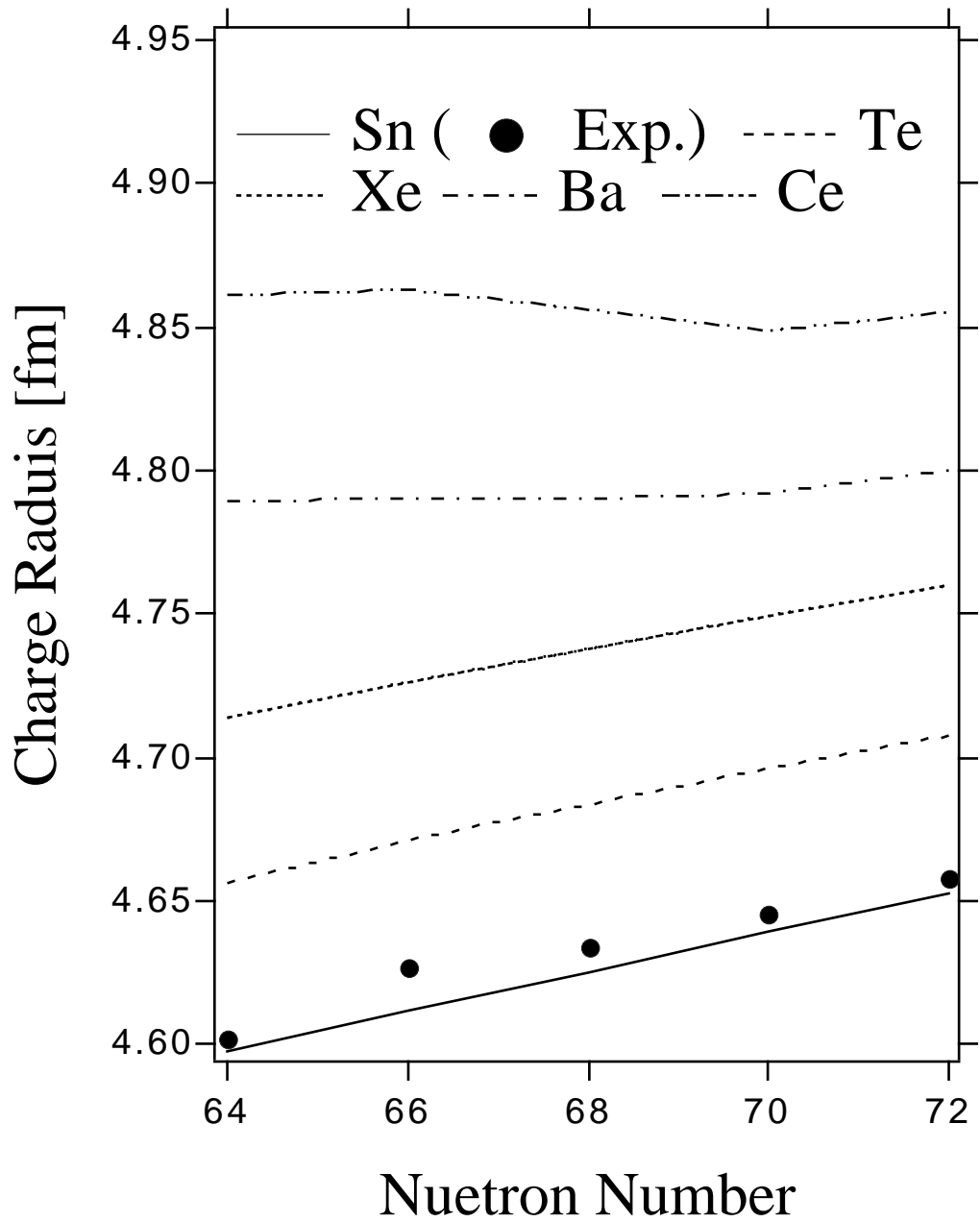
\item[Fig.7]
The charge radius calculated using the RMF theory with axial deformation is
shown as a function of the neutron number.
The experimental data are extracted from Ref.~\cite{DeV87}.






\end{description}











\begin{thebibliography}{999}

\bibitem{Tan85}	I. Tanihata, H. Hamagaki, O. Hashimoto, Y. Shida, N.
Yoshikawa, K.
Sugimoto, O. Yamakawa, T. Kobayashi and N. Takahashi, Phys. Rev. Lett. {\bf
55}
(1985) 2676.
\bibitem{Tan92}	I. Tanihata, D. Hirata, T. Kobayashi, S. Shimoura, K.
Sugimoto and
H. Toki, Phys. Lett. {\bf B289} (1992) 261.
\bibitem{Suz95} T. Suzuki, H. Geissel, O. Bochkarev, L. Chulkov,
M. Golovkov, D. Hirata, H. Irnich, Z. Janas, H. Keller, T. Kobayashi,
G. Kraus, G. M\"unzenberg, S. Neumaier, F. Nickel, A. Ozawa,
A. Piechaczeck, E. Roeckl, W. Schwab, K. S\"ummerer, K. Yoshida,
I. Tanihata, Phys. Rev. Lett. {\bf 75} (1995) 3241.
\bibitem{Ham89}   J. H. Hamilton, Treatise on Heavy Ion Science, ed. D. A.
Bromley
(Plenum, New York), {\bf 8} (1989) 3.
\bibitem{Ott89}   E. W. Otten, Treatise on Heavy Ion Science, ed. D. A.
Bromley
(Plenum, New York), {\bf 8} (1989) 517.
\bibitem{Tan95}  I.Tanihata, Progress in Particle and Nuclear Physics
                {\bf 35} (1995) 505
\bibitem{RNB4} See for example, Proceedings of the fourth International
Conference on Radioactive Nuclear Beams, Ohmiya, Japan, 1996,
Nucl. Phys. {\bf A616} (1997).
\bibitem{Ser86}	B. D. Serot and J. D. Walecka, Adv. Nucl. Phys.
{\bf 16}
(1986) 1.
\bibitem{Ser92} B. D. Serot, Rep. Prog. Phys. {\bf 55} (1992) 1855.
\bibitem{Bro90}	R. Brockmann and R. Machleidt, Phys. Rev. {\bf C42}
(1990)
1965.
\bibitem{Eng94} L. Engvik, M. Hjorth-Jensen, E. Osnes, G. Bao
and E. Ostgaard, Phys. Rev. Lett. {\bf 73} (1994) 2650.
\bibitem{Bro92} R. Brockmann and H. Toki, Phys. Rev. Lett.{\bf B68} (1992)
3408.
\bibitem{Rei86}	P.-G. Reinhard, M. Rufa, J. Maruhn, W. Greiner and J.
Friedrich, Z.
Phys. {\bf A323} (1986) 13.
\bibitem{Gam90}	Y. K. Gambhir, P. Ring and A. Thimet, Ann. of Phys.
{\bf
198} (1990)
132.
\bibitem{Hir91}	D. Hirata, H. Toki, T. Watabe, I. Tanihata and B. V.
Carlson, Phys.
Rev. {\bf C44} (1991) 1467.
\bibitem{Sum93}	K. Sumiyoshi, D. Hirata, H. Toki and H. Sagawa,
Nucl. Phys.
{\bf A552} (1993) 437.
\bibitem{Pom97} K. Pomorski, P. Ring, G.A. Lalazissis,
A. Baran, Z. Lojewski, B.Nerlo--Pomorska and M. Warda,
Nucl. Phys. {\bf A624} (1997) 349.
\bibitem{Sha92}	M. M. Sharma and P. Ring, Phys. Rev. {\bf C46}
(1992) 1715.
\bibitem{Hir93}	D. Hirata, H. Toki, I. Tanihata and P. Ring, Phys.
Lett.
{\bf B314} (1993) 168.
\bibitem{Lal95}	G. A. Lalazissis and M. M. Sharma, Nucl. Phys. {\bf
A586}
(1995)
201.
\bibitem{Shen98}H. Shen,  H. Toki, K. Oyamatsu and K. Sumiyoshi,
Prog. Theor. Phys. {\bf 100} (1998) 1013.
\bibitem{Hir97} D. Hirata, K. Sumiyoshi, I. Tanihata, Y. Sugahara,
T. Tachibana, and H. Toki, Nucl. Phys. {\bf A616} (1997) 438;
RIKEN Preprint No. RIKEN-AF-NF-268
\bibitem{Hir96}	D. Hirata, K. Sumiyoshi, B. V. Carlson, H. Toki and I.
Tanihata, Nucl. Phys. {\bf A609} (1996) 131.
\bibitem{Hir95} D. Hirata, H. Toki and I. Tanihata,
Nucl. Phys. {\bf A589} (1995) 239.
%
\bibitem{Cas85} R. F. Casten, P. von Brentano, K. Heyde,
P. Van Isacker and J. Jolie, Nucl. Phys. {\bf A439} (1985) 289.
\bibitem{Wys89} R. Wyss, A. Granderath, W. Lieberz, R. Bengtsson,
P. von Brentano, A. Dewald, A. Gelberg, A. Gizon, J. Gizon,
S. Harrisopulos, A. Johnson, W. Nazarewicz, J. Nyberg
and K. Schiffer, Nucl. Phys. {\bf A505} (1989) 337.
\bibitem{Gel93} A. Gelberg, D. Lieberz, P. von Brentano,
I. Ragnarsson, P. B. Semmes and I. Wiedenh\"over,
Nucl. Phys. {\bf A557} (1993) 439c.
\bibitem{Mey97} U. Meyer, A. Faessler and S. B. Khadkikar,
Nucl. Phys. {\bf A624} (1997) 391.
%
\bibitem{Dew92} A. Dewald, P. Sala, R. Wrzal, G. B\"ohm,
D. Liebrez, G. Siems, R. Wirowski, K. O. Zell, A. Gelberg,
P. von Brentano, P. Nolan, A. J. Kirwan, P. J. Bishop,
R. Julin, A. Lampinen and J. Hattula, Nucl. Phys.
{\bf A545} (1992) 822.
\bibitem{Pet94} P. Petkov, R. Kr\"ucken, A. Dewald, P. Sala,
G. B\"ohm, J. Altmann, A. Gelberg, P. von Brentano, R.V. Jolos
and W. Andrejtscheff, Nucl. Phys. {\bf A568} (1994) 572.
%
\bibitem{Bon85} P. Bonche, H. Flocard, P. H. Heenen, S. J. Krieger
and M. S. Weiss, Nucl. Phys. {\bf A443} (1985) 39.
\bibitem{Koe88} W. Koepf and P. Ring, Phys. Lett. {\bf B212} (1988) 397.
\bibitem{Abe90} S. Aberg, H. Flocard and W. Nazarewicz,
Annu. Rev. Nucl. Part. Sci. {\bf 40} (1990) 439
and references therein.
\bibitem{Wer94} T. R. Werner, J. A. Sheikh, W. Nazarewicz,
M. R. Strayer, A. S. Umar and M. Misu,
Phys. Lett. {\bf B333} (1994) 303.
%
\bibitem{Sug94}	Y. Sugahara and H. Toki, Nucl. Phys. {\bf A579}
(1994) 557.
\bibitem{Bog77}	J. Boguta and A. R. Bodmer, Nucl. Phys. {\bf A292}
(1977) 413.
\bibitem{Bod91}	A. R. Bodmer, Nucl. Phys. {\bf A526} (1991) 703.
\bibitem{Sug95}	Y. Sugahara, Doctor thesis, Tokyo Metropolitan
University
(1995).
\bibitem{Tok95} H. Toki, D. Hirata, Y. Sugahara, K. Sumiyoshi and I. Tanihata,
Nucl. Phys. {\bf A588} (1995) 357c.
\bibitem{Flo73}	H. Flocard, P. Quentin, A. K. Kerman and D.
Vautherin, Nucl.
Phys.
{\bf A203} (1973) 433.
\bibitem{Taj96}	N. Tajima, S. Takahara and N. Onishi,
Nucl. Phys. {\bf A603} (1996) 23.
\bibitem{Pair} L. S. Kisslinger and R. A. Sorensen, Mat. Fys. Medd. Dan. Vid. Selsk.
{\bf 32} No. 9 (1960) 5.
\bibitem{Men96} J. Meng and P. Ring,
Phys. Rev. Lett. {\bf 77} (1996) 3963.
\bibitem{Bre98} D. Hirata and B. V. Carlson, ENAM98, ed. B. M. Sherrill, D. J. Morrissey and C. N. Davids
(The American Institue of Physics) (1998) 527.
 \bibitem{Ram87}	S. Raman, C. H. Malarkey, W. T. Milner, C. W.
Nestor, Jr., and P.
 H. Stelson, At. Data Nucl. Data Tables {\bf 36} (1987) 1.
\bibitem{Aud95}	G. Audi and A.H. Wapstra, Nucl. Phys. {\bf A595}
(1995) 409.
 \bibitem{DeV87}	H. De Vries, C. W. De Jager and C. De Vries,
 At. Data Nucl. Data Tables {\bf 36} (1987) 495.
\end{thebibliography}
\end{document}